\documentclass[prl,a4paper,twocolumn,showpacs]{revtex4}
\usepackage{graphicx}

\begin{document}

\title{Ultrafast interatomic electronic decay in multiply excited clusters}

\author{Alexander I. \surname{Kuleff}}
\email[E-mail: ]{alexander.kuleff@pci.uni-heidelberg.de}
%\altaffiliation[On leave from: ]{INRNE, BAS, 72, Tzarigradsko Chaussee Blvd., 1784 Sofia, Bulgaria}
\author{Kirill \surname{Gokhberg}}
%\email[E-mail: ]{kirill.gokhberg@pci.uni-heidelberg.de}
\author{S\"oren \surname{Kopelke}}
%\email[E-mail: ]{kirill.gokhberg@pci.uni-heidelberg.de}
\author{Lorenz S. \surname{Cederbaum}}
%\email[E-mail: ]{lorenz.cederbaum@pci.uni-heidelberg.de}
%\\
\affiliation{Theoretische Chemie, PCI, Universit\"at Heidelberg\\
             Im Neuenheimer Feld 229, 69120 Heidelberg, Germany}

\date{\today}

\begin{abstract}
An ultrafast mechanism belonging to the family of interatomic Coulombic decay (ICD) phenomena is proposed. When two excited species are present, an ultrafast energy transfer can take place bringing one of them to its ground state and ionizing the other one. It is shown that if large homoatomic clusters are exposed to an ultrashort and intense laser pulse whose photon energy is in resonance with an excitation transition of the cluster constituents, the large majority of ions will be produced by this ICD mechanism rather than by two-photon ionization. A related collective-ICD process that is operative in heteroatomic systems is also discussed.
\end{abstract}


\maketitle

The rapid development during the last decades of very intense light sources with extreme short pulse duration opened a new era in the study of radiation-matter interaction. Studying the interaction of intense fields with matter brought to the discovery of a whole plethora of new physical phenomena, like high-harmonic generation, above-threshold ionization, or tunneling ionization, to name only a few. In the same time, the progress in generating extremely short pulses gave the scientific community a powerful tool to monitor and control the electron dynamics in atomic and molecular systems and to study processes that take place on a time scale in which the electronic motion is still disentangled from the slower nuclear dynamics (for recent reviews see, e.g., Refs.~\cite{Misha_atto,Giuseppe_atto}). A number of free-electron lasers are in operation today providing extremely bright, coherent, and ultrashort pulses in the VUV regime. Exposed to such highly intense pulses, atomic and molecular systems will absorb a large amount of photons triggering various dynamical effects. In this letter we will restrict ourselves to situations where the single-photon energy in the pulse is not high enough to directly ionize the system. It is well known that even in this case the system can be ionized by a multiphoton ionization mechanism. The multiphoton ionization (MPI) results from the ability of quantum systems to absorb several and even many photons, whose individual energies are insufficient to ionize the system. The combined energy of the absorbed photons, though, suffices to eventually eject one or many electrons from the system. During the last decade the MPI has been intensively studied also in composite systems, like clusters, employing the new powerful laser sources (for a review see, e.g. Ref. \cite{Rost_rev06}). However, little attention was paid to other mechanisms that can lead to a multiple ionization in an atomic or molecular cluster irradiated by an intense laser pulse.

\begin{figure}[ht] 
\begin{center}
\includegraphics[width=6cm]{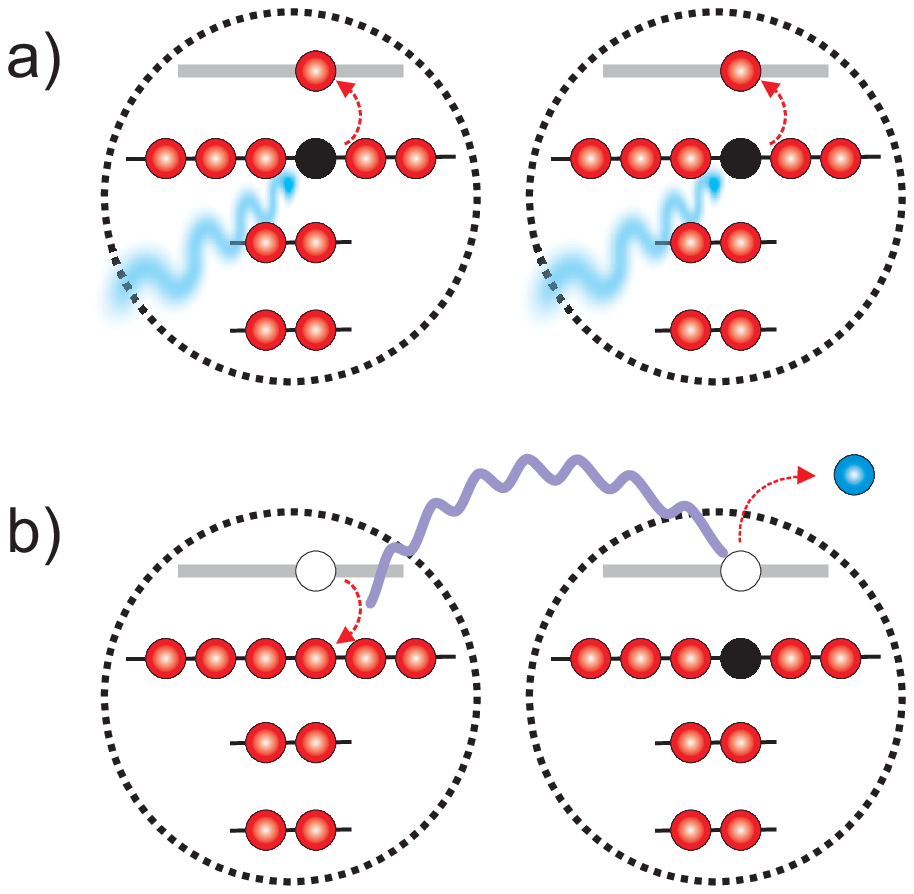}
\end{center}
\caption{\label{scheme}Schematic representation of the process. a) Two subunits of the system are excited by absorbing two photons. b) One of the constituents of the system de-excites transferring the energy to the neighbor which uses it to emit its excited electron.}
\end{figure}

In this letter we aim at discussing a hitherto unrecognized mechanism for producing ionized species in homoatomic or homomolecular clusters exposed to an intense laser pulse, which in many cases can be by far the dominating one. For simplicity we will consider an atomic cluster, but we stress that the process is general and not restricted to atomic systems. Let us take a homoatomic cluster and irradiate it with a short and intense laser pulse with photon energy below the ionization threshold of the cluster constituents but in resonance with one of their excited states. A fraction of the cluster constituents will be ionized by MPI but, since we are at resonance, the large majority of atoms will be exited. The well known effect of Coulomb blockade will not play a significant role here, since we suppose that the system is exposed to a very short pulse, i.e. to a broadband excitation. Thus, there will be many excited atoms in the cluster whose neighbor is also excited. Having two excited atoms in close proxmity the following interatomic electronic decay mechanism is conceivable. One of the atoms is de-excited, the energy is transferred to the other one which uses it to emit its excited electron. Thus, at the end of the process, one of the atoms is ionized and the other one has returned to its ground state. The process is pictorially represented in Fig.~\ref{scheme} and can be written in short as:
\[
A^\ast\cdots A^\ast \rightarrow A\cdots A^+ + e^-.
\]

This process bears similarities with the interatomic (intermolecular) Coulombic decay (ICD) predicted theoretically more than ten years ago \cite{ICD1} and since then studied very intensively both theoretically and experimentally (see, e.g. Refs.~\cite{Robin_PR,Vitali_rev,Uwe1,Reinhard1,Reinhard_water,Uwe_water}). The ICD is a very efficient electronic decay mode of inner-valence ionized atoms or molecules embedded in an environment. Inner-valence ionized states usually have energies below the double ionization threshold and, thus, cannot autoionize. However, here the environment plays a critical role. When the initially ionized atom or molecule has neighbors, like in a cluster, an electron from a higher level may fill the vacancy and the released energy can be transferred to a neighbor form which a secondary electron is emitted. Thus, the creation of a single hole in one of the subunits in the system leads to the formation of two positively charged subunits that repel each other typically leading to a Coulomb explosion that disintegrates the system. The process is ultrafast with typical lifetimes of few to few tens of femtoseconds, quenching all other energetically allowed relaxation modes of the system. The discovery of the ICD revealed a whole zoo of related phenomena, involving both energy and electron transfer and initiated by single or multiple ionization, as well as by inner- or outer valence excitation (for recent review, see Ref.~\cite{Vitali_rev}). Although these processes have different names and acronyms, we will refer here to all these phenomena as ICD in order to make the text more transparent. The only, but important difference of the process proposed here and the ICD phenomena studied until now is that the ICD assumes an excited system interacting with a non-excited environment, while in the process sketched in Fig.~\ref{scheme} the distinction between system and environment is not possible. On the contrary, both constituents are equally suitable to undergo an electronic decay. However, we will refrain from giving a new name to the process discussed in this letter and will refer to it as ICD.

The important question is, of course, whether this ICD process is efficient enough and can compete with the other possible de-excitation modes (e.g. photon emission) in the dimer or, even more interesting, in a large cluster. To estimate that we have to calculate the rate of the process, or the decay width $\Gamma$. The easiest way to estimate the decay width is to consider the process within the simplified but insightful picture of interaction between two dipoles via a virtual-photon exchange. A virtual photon is emitted as a result of the de-excitation of one of the excited atoms and then absorbed by the other excited atom causing its ionization. The virtual photon exchange picture, which is correct at large interatomic distances, enables the derivation of analytical formulae for the decay width \cite{Vitali-PRL04}. Such formulae exhibit $1/R^6$ dependence ($R$ being the interatomic separation) with a prefactor specific for the emitting and absorbing constituents and accounting for the dipole selection rules of the involved transitions. The derivation of such an expression is straightforward using the procedure explained in detail in Ref.~\cite{Kirill-asympt}. 

In order to illustrate the efficiency of the ICD mechanism proposed here we consider a concrete example. Let us take a neon dimer in which both of the neon atoms are in their first excited state Ne$^\ast (2p^{-1}3s)$-Ne$^\ast (2p^{-1}3s)$. The energy of Ne$^\ast (2p^{-1}3s)$ is about 16.7~eV above the ground state while the ionization potential (IP) of the neon atom is about 21.6~eV. Thus, the $3s\to 2p$ transition in one of the neons will release enough energy to ionize the other one, emitting an electron with kinetic energy of about 11.8~eV. Averaging over the multiplicities of the initial states and summing over the final states we obtain for the total decay width of the system Ne$^\ast(2p^{-1}3s)$-Ne$^\ast(2p^{-1}3s)$ as a function of the internuclear distance $R$ the following expression (in atomic units)
\begin{equation}\label{gamma}
 \Gamma(R)=\frac{3c\,f\sigma}{\pi\omega^2}\frac{1}{R^6},
\end{equation}
where $f$ is the oscillator strength of the $3s\rightarrow 2p$ transition, $\sigma$ is the ionization cross section of Ne$^\ast(2p^{-1}3s)$, $c$ is the speed of light, and $\omega$ is the virtual-photon energy.

The values of the quantities entering Eq.~(\ref{gamma}) are known from the literature -- the oscillator strength for the $3s\to 2p$ transition in neon is 0.16 \cite{Brion} and the photoionization cross section of Ne$^\ast(2p^{-1}3s)$ is about 0.18~Mb \cite{Kau_etal96}. At the Ne$_2$ equilibrium distance of 3.1~\AA{} the decay width for the process is 0.24~meV, which implies a life time of about 2.8~ps. This is 3 orders of magnitude faster than photon emission, which is known to be about 2~ns \cite{ne_radltime}, and thus ICD is by far the dominant relaxation pathway in the dimer. The results obtaind by the virtual-photon model are correct at large interatomic distances $R$. They are very promissing, in particular, since it is known from previous studies \cite{Vitali-PRL04} that such kind of asymptotic formulae underestimate the decay rates around equilibrium distances due to neglecting the orbital overlap. 

\begin{figure}[ht] 
\begin{center}
\includegraphics[width=6cm,angle=270]{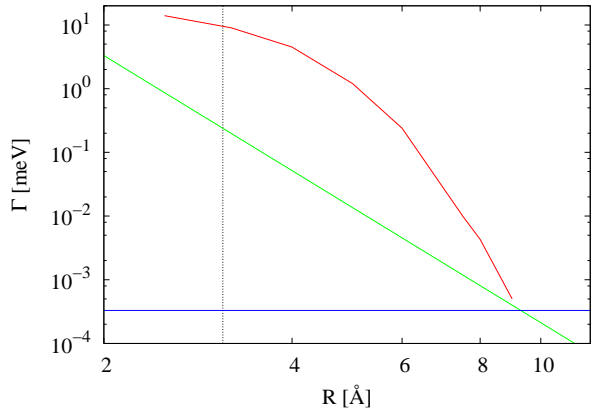}
\end{center}
\caption{\label{width}Total ICD width $\Gamma$ for the system Ne$^\ast(2p^{-1}3s)$-Ne$^\ast(2p^{-1}3s)$ compared to the prediction of virtual photon model, Eq.~(\ref{gamma}). The atomic fluorescence decay width is indicated by a horisontal line, while the equilibrium interatomic separation by a vertical one. Note the double logarithmic scale used.}
\end{figure}

In order to have more reliable values for the decay rate, we used the ${\cal L}^2$ \textit{ab initio} method, known as Fano-Stieltjes approach \cite{Fano-Stieltjes-ADC}. In this method the boundlike and the continuumlike components of the wave function of the decaying state are constructed using the Green's function formalism, and the problem of the normalization of the continuum wave function is addressed by using the Stieltjes imaging technique (see Ref. \cite{Fano-Stieltjes-ADC} for details). The \textit{ab initio} results are shown in Fig.~\ref{width} together with the predictions of the virtual-photon model, Eq.~(\ref{gamma}). For a reference, the atomic fluorescence decay width is also shown in the figure. We see that up to about 9~\AA{} of internuclear separation, i.e. about 3 times the equilibrium distance, the asymptotic formula largely underestimates the ICD decay width. When the width becomes very small (i.e., at large internuclear separation) the \textit{ab initio} method suffers fron numerical instabilities and cannot be safely employed. It is also around 9~\AA{} distance where the radiative decay becomes competitive. At the equilibrium distance of the neon dimer the \textit{ab initio} computation predicts a total decay width of 5.4~meV which is more than 20 times larger than the virtual photon result. This decay width corresponds to a life time as short as 122~fs, which means that the ICD sets in before the nuclear dynamics play a role. 

Let us now comment on larger clusters. Most importantly, since the total decay width is a sum of the partial widths of all possible decay channels, it is clear that if we have more than two interacting excited atoms the ICD process will become dramatically faster \cite{Robin_PR,Bjoernholm_PRL2004}. In (Ne$^\ast$)$_4$, for example, there are 12 open channels, which suggests that the ICD life time in this cluster will be 6 times shorter than that for (Ne$^\ast$)$_2$. Thus, in big clusters, where a resonant intense laser pulse will produce a large number of excited atoms, the ICD mechanism will be extremely efficient.

Once we have seen that the ICD process is ultrafast and can be expected to outperform other possible ways of relaxation, let us return to the question of the competition between the ICD and the MPI in the production of positive ions in a cluster irradiated by a laser pulse with high density of photons. For that purpose, it is illuminating to consider again a concrete example. Let a Ne$_{1000}$ cluster be exposed to a short and intense laser pulse with photon energy of 16.7~eV, i.e. resonant to the $2p\to 3s$ excitation of the neon atom. We can estimate the number of excited atoms and the number of those ionized by two-photon ionization in the cluster after the pulse by solving the following system of rate equations
\begin{eqnarray}\label{rates}
 &&\frac{dN(t)}{dt}=-\sigma_0\Phi(t)N(t) - \sigma_2\Phi^2(t)N(t),  \nonumber \\
 &&\frac{dN^{(\ast)}(t)}{dt}=\sigma_0\Phi(t)N(t) - \sigma_1\Phi(t)N^{(\ast)}(t), \nonumber \\
 &&\frac{dN^{(+)}(t)}{dt}=\sigma_1\Phi(t)N^{(\ast)}(t) + \sigma_2\Phi^2(t)N(t). 
\end{eqnarray}
In Eqs.~(\ref{rates}) $N(t)$, $N^{(\ast)}(t)$, and $N^{(+)}(t)$ are the number of neutral, excited, and ionized by two-photon ionization atoms as a function of time, respectively, while  $\sigma_0$ is the absorption cross section, $\sigma_1$ is the photoionization cross section of Ne$^\ast(2p^{-1}3s)$, $\sigma_2$ is the two-photon ionization cross section, and $\Phi(t)$ denotes the photon flux which contains the information on the temporal profile of the pulse. In order to obtain quantitative results, one has to consider also the spatial profile of the pulse and the geometry of the irradiated cluster. However, we aim here at making only an estimate of the ratio between $N^{(\ast)}$ and $N^{(+)}$ after the pulse and that is why we will use the rather simplified picture of a rectangular pulse with intensity 10$^{12}$~W/cm$^2$ and duration 50~fs, ignoring the dependence of the laser-cluster interaction on the spatial profile of the pulse and the geometry of the cluster. In this case, Eqs.~(\ref{rates}) can be easily solved and using the atomic data, $\sigma_0\approx 273$~Mb \cite{Brion}, $\sigma_1\approx 0.18$~Mb \cite{Kau_etal96}, and $\sigma_2\approx 2\times 10^{-49}$~cm$^4$s \cite{Ne_MPI}, one obtains that in the Ne$_{1000}$ cluster after the pulse 991 atoms will be excited and only 3 will be ionized by a two-photon ionization. Since, as we saw, the ICD process is very efficient, one would expect that every pair of Ne$^\ast(2p^{-1}3s)$ will undergo ICD producing about 495 neon ions. Thus, the ratio of the neon ions produced by ICD and those produced by a two-photon ionization is about 166:1. It is clear that by increasing the laser intensity one will produce more ions by two-photon ionization, while decreasing it will favor the ion production via ICD mechanism. For example, with a laser intensity of 10$^{13}$~W/cm$^2$ this ratio is 13.7:1. We see that even at these relatively high intensities, the ICD mechanism is still by far the dominant source of ionized species in the cluster. It is clear that even at higher intensities, the production of ions via ICD has to be taken into account when interpreting experimental results. Indeed, the peak intensity is achieved only in the focal point of the laser which usually is much smaller than the interaction region. A large fraction of the clusters, thereby, will be exposed to a less intense field where the ICD is the dominant ion-production source.

\begin{figure}[ht] 
\begin{center}
\includegraphics[width=6cm]{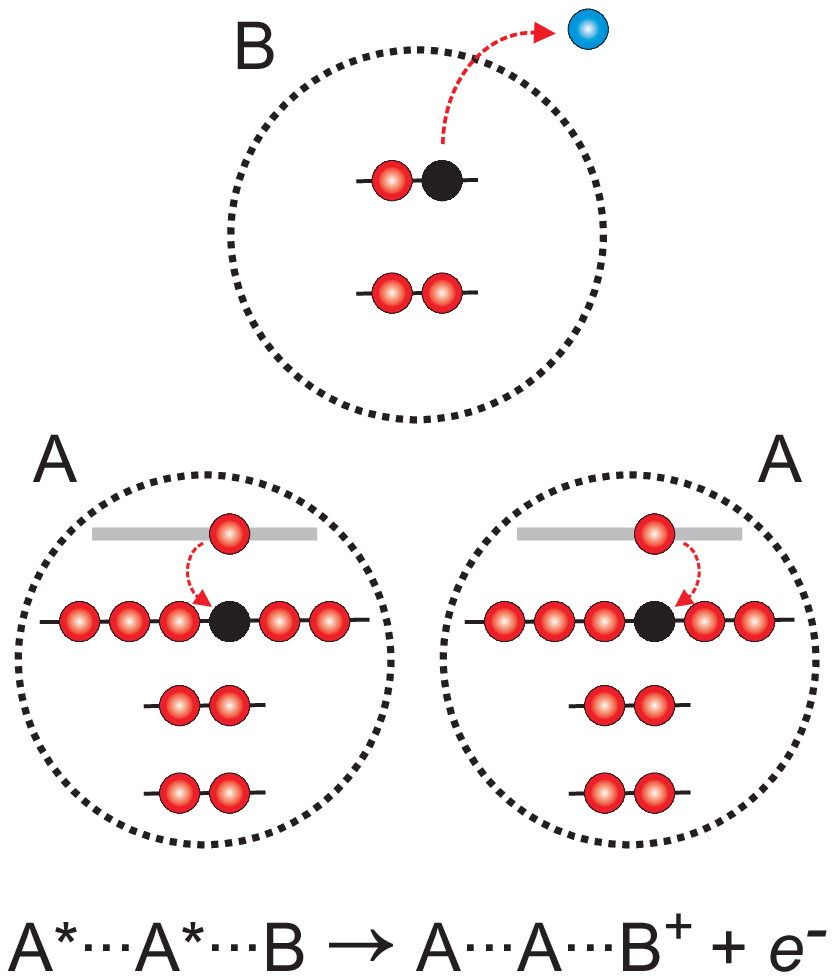}
\end{center}
\caption{\label{cETI}Schematic representation of the collective process discussed in the text for heteroatomic clusters. Two constituents of the system de-excite simultaneously transferring the energy to a third species and ionizing it.}
\end{figure}

At the end we would like to comment briefly on another possibility to create ionized species in multiply excited clusters which will be operative in the case of heteroatomic systems. In the case when the de-excitation energy of an excited atom is insufficient to ionize another atom, a process related to the recently discussed collective-ICD \cite{Vitali-cICD} can take place. In the collective-ICD process two inner-valence ionized species de-excite simultaneously transferring their ``collective'' energy to a third neighbor and ionizing it. In analogy, one can think about a collective-ICD where two excited atoms or molecules de-excite simultaneously and the released energy is used by a third atom or molecule to eject one of its electrons, see Fig. \ref{cETI}. An important point to note is that, in contrast to the former case, in the case of collective-ICD from excited species the process will not have to compete with the Coulomb explosion dynamics of the two neighboring ions. It is clear that the collective-ICD from excited species will be energetically open when $2E(A^\ast)>IP(B)$. This implies that $A$ should be different from $B$ since we supposed that the ICD process of Fig. \ref{scheme} is energetically closed. Of course, if the ICD channel is open, the collective decay can also take place, but since it involves three electrons, its importance compared to the two-electron ICD process will be low.

Let us conclude. In this letter we proposed a hitherto unrecognized mechanism for producing ionized species in multiply excited atomic or molecular clusters. The mechanism belongs to the family of interatomic (intermolecular) Coulombic decay phenomena and consists of an ultrafast energy transfer between two excited species, bringing one of them to its ground state and ionizing the other. We showed that the process is ultrafast (in the femtosecond time regime) and as such is extremely efficient compared to other possible relaxation modes. Moreover, we showed that if large clusters are exposed to an ultrashort and intensive laser pulse ($10^{12}-10^{13}$ W/cm$^2$ in the present example) which is in resonance with an excitation transition of the cluster constituents, the large majority of ions will be produced by this ICD mechanism rather than by two-photon ionization. In addition, we proposed a collective ICD process that can take place in heteroatomic or heteromolecular systems also yielding ionized species. We hope that our work will trigger more theoretical and experimental investigations of these ICD effects in systems exposed to ultrafast laser pulses with high density of photons.

The authors thank K. Ueda for stimulating discussions and for sharing with us his experimental data prior to 
publication which triggered the present work. The research leading to these results has received funding from the European Research Council under the European Community's Seventh Framework Programme (FP7/2007-2013) /
ERC Advanced Investigator Grant n$^\circ$ 227597.

\end{document}